\documentclass[aip,rsi,reprint,graphicx]{revtex4-1} 

\usepackage{graphicx}
\usepackage{gensymb}
\usepackage{epstopdf}

\begin{document}


\title{The Suborbital Particle Aggregation and Collision Experiment (SPACE): Studying the Collision Behavior of Submillimeter-Sized Dust Aggregates on the Suborbital Rocket Flight REXUS 12} 

\author{Julie Brisset}
\email{j.brisset@tu-bs.de}
\author{Daniel Hei{\ss}elmann}
\affiliation{Institut f{\"{u}}r Geophysik und extraterrestrische Physik (IGeP), TU Braunschweig, Mendelssohnstr. 3, 38102 Braunschweig, Germany \\
           Phone: +49-531-3917236\\
           Fax: +49-531-3918126}
\affiliation{Max Planck Institute for Solar System Research, Max-Planck Str. 2, 37191 Katlenburg-Lindau, Germany}
\author{Stefan Kothe}
\author{Ren{\'{e}} Weidling}
\author{J{\"{u}}rgen Blum}

\affiliation{Institut f{\"{u}}r Geophysik und extraterrestrische Physik (IGeP), TU Braunschweig, Mendelssohnstr. 3, 38102 Braunschweig, Germany \\
           Phone: +49-531-3917236\\
           Fax: +49-531-3918126}

\date{\today}

\begin{abstract}

The Suborbital Particle Aggregation and Collision Experiment (SPACE) is a novel approach to study the collision properties of submillimeter-sized, highly porous dust aggregates. The experiment was designed, built and carried out to increase our knowledge about the processes dominating the first phase of planet formation. During this phase, the growth of planetary precursors occurs by agglomeration of micrometer-sized dust grains into aggregates of at least millimeters to centimeters in size. However, the formation of larger bodies from the so-formed building blocks is not yet fully understood. Recent numerical models on dust growth lack a particular support by experimental studies in the size range of submillimeters, because these particles are predicted to collide at very gentle relative velocities of below 1 cm/s that can only be achieved in a reduced-gravity environment.\\

The SPACE experiment investigates the collision behavior of an ensemble of silicate-dust aggregates inside several evacuated glass containers which are being agitated by a shaker to induce the desired collisions at chosen velocities. The dust aggregates are being observed by a high-speed camera, allowing for the determination of the collision properties of the protoplanetary dust analog material. The data obtained from the suborbital flight with the REXUS (Rocket Experiments for University Students) 12 rocket will be directly implemented into a state-of-the-art dust growth and collision model.

\end{abstract}


\maketitle 

\section{Introduction}
\label{s:intro}

Planet formation is an active field of observational, experimental and theoretical astrophysical research. Within the last two decades more than 800 planets have been found orbiting around other stars, which proves that the formation of planets is not a process restricted to our own Solar System. Today we know that the kilometer-sized precursors of planets, the so-called \emph{planetesimals}, form in protoplanetary disks (PPDs), which are collapsed molecular clouds of gas and dust around young stars. The formation of larger bodies starts with the coagulation of colliding (sub-)micrometer-sized dust grains. However, due to the distance and the opacity of these disks, as well as due to the astronomical timescales involved (thousands to millions of years), direct observation of the growth processes is not possible. Therefore, theoretical investigations, backed up by experiments, are needed to understand the first phases of planet formation.

The collisions within the initial population of grains are driven by the interaction between the dust and the gas of the PPD \cite{weidenschilling1977MNRAS}. For small particles (with typical sizes of a few $\mu$m), the collision velocities are so low that every collision leads to sticking between the dust particles \cite{blum_wurm2008ARA&A}. As typically the collision speed increases with increasing dust-aggregate size\cite{weidenschilling1977MNRAS}, the outcome of these collisions is not easily predictable. Previous laboratory work has shown that while at low velocities sticking occurs, higher velocities lead to bouncing of the aggregates or even result in their fragmentation\cite{guettler_et_al2010A&A}. The two latter effects obviously constrain the growth of larger dust agglomerates and it is, thus, unclear how kilometer-sized \emph{planetesimals} form.

\begin{figure}[!bth]
\includegraphics[width=0.45\textwidth]{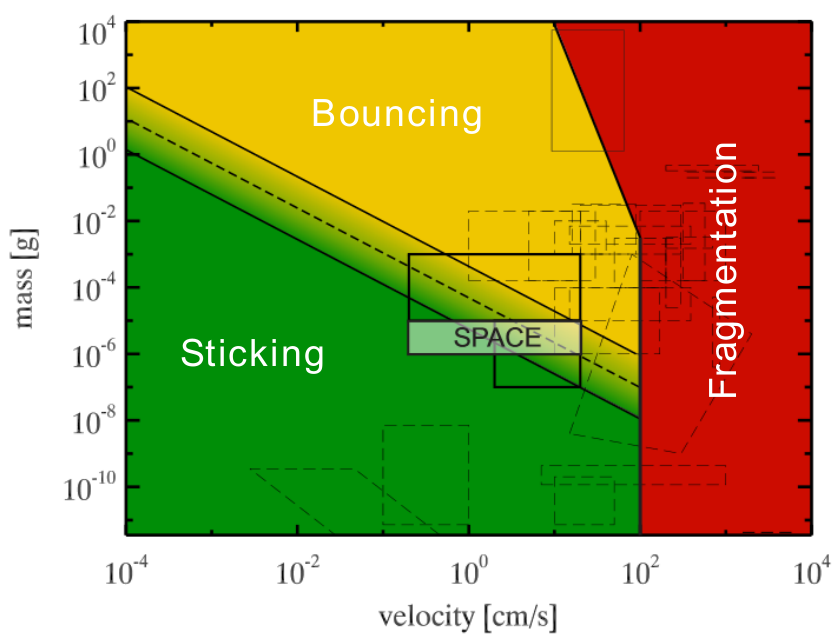}
\caption{Mass-velocity plot for equal-sized protoplanetary dust aggregates after \citet{guettler_et_al2010A&A}. The colors denote typical model predictions for sticking (green), bouncing (yellow), and fragmentation (red), respectively. The parameter space so far covered by dust-collision experiments are shown by the dotted boxes. The white box indicates the parameter space that can be covered by the SPACE experiment.}
\label{f:ghanaplot}
\end{figure}

\citet{guettler_et_al2010A&A} unified experimental and theoretical work to understand the growth of dust agglomerates. 19 laboratory experiments on dust-aggregate collisions were reviewed and compiled into a detailed collision model that predicts the outcome of collisions between dust aggregates of all masses and collision velocities (see Figure \ref{f:ghanaplot} for an example of the collision model). This model was used by \citet{zsom_et_al2010A&A} as input for a Monte Carlo growth simulation, which describes the evolution of dust particles in protoplanetary disks. These numerical simulations have shown that the particle growth stops at sizes on the order of a few millimeters, since collisions of these and of larger particles lead to bouncing (yellow region if Figure \ref{f:ghanaplot}).

However, the result of the simulations strongly depends on the transitions between the different collision outcomes in the input model. \citet{guettler_et_al2010A&A} extrapolated the transition between the different outcomes over a wide range of the parameter space, because of a lack of experimental data in these regions. Therefore, several new experiments have been designed to investigate the regions of the parameter space that is important for the growth simulations. These studies particularly focus on the transition between sticking and bouncing. Because this transition is predicted by the model to occur at collision velocities lower than 1 cm/s, it is not feasible to perform the experiments in the laboratory. Additionally, a large number of collisions have to be observed to make a reliable conclusion about the transition regime. This can be performed either by conducting several short (up to 9 s of microgravity) experiments in a drop tower (\citet{kothe_et_al2013Icarus} and \citet{weidling_et_al2012Icarus}) or longer (up to 180 s of microgravity) experiments on a suborbital rocket.

One of the new experimental investigations is the Suborbital Particle Aggregation and Collision Experiment (SPACE) that flew onboard the REXUS 12 rocket in March 2012. The aim of the SPACE experiment was to study collisions within an ensemble of submillimeter-sized particles under microgravity conditions at velocities relevant to PPDs.

\section{Experimental Setup}
\label{s:setup}
\subsection{The REXUS Rocket}
\label{s:REXUS}
REXUS stands for Rocket Experiments for University Students. The REXUS/BEXUS (the associated atmospheric balloon) program is a combined project shared by the German aerospace agency DLR and the Swedish National Space Board (SNSB), launching two ballistic rockets per year. The launch site is at Esrange near Kiruna in northern Sweden.

\begin{figure*}[!bth]
\includegraphics[width=\textwidth]{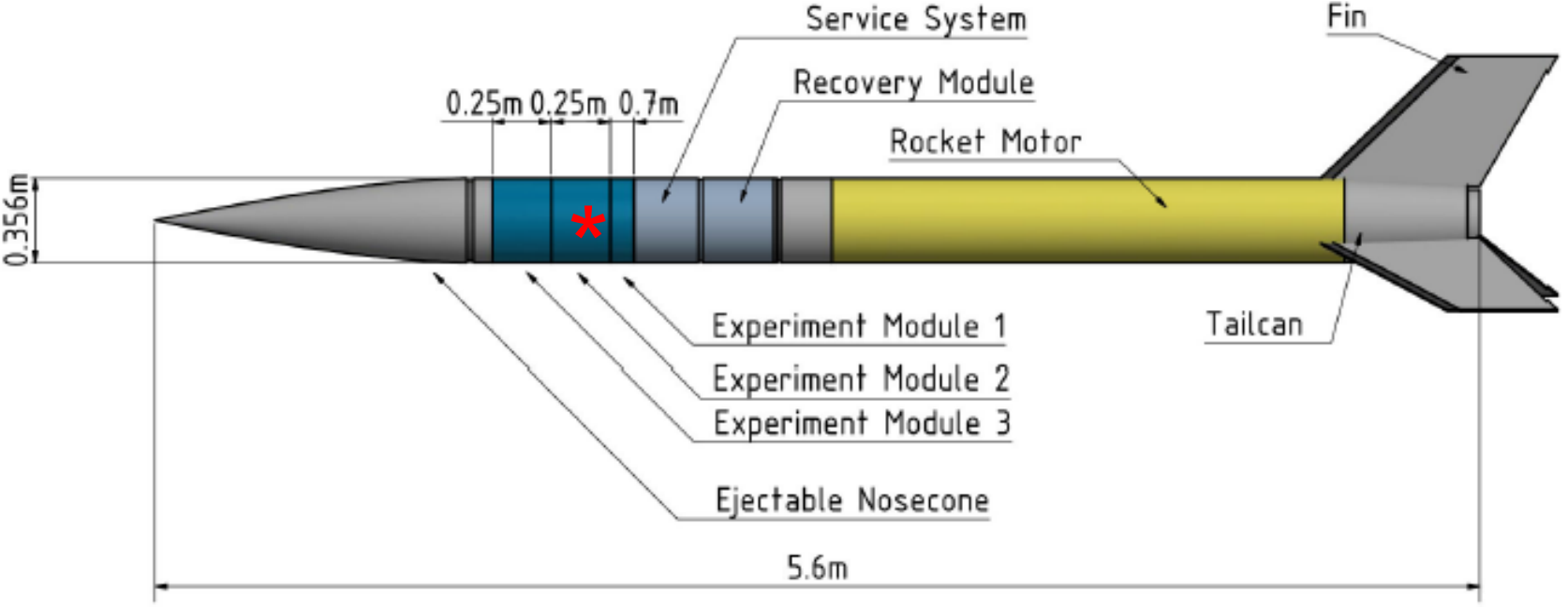} 
\caption{A typical REXUS configuration, with 3 experiments aboard (one in the nosecone) \cite{rexus_manual}. The position of the SPACE experiment within the rocket is marked by the red asterisk.}
\label{f:rexus}
\end{figure*}

The REXUS rocket (Figure \ref{f:rexus}) is 5.6 m long and has a diameter of 356 mm. It can host 3 to 5 experiments depending on their dimensions and mass; the maximum total payload mass is about 95 kg. The rocket is composed of an improved Orion motor, a recovery module containing a parachute and a GPS system, a service module supplying the experiments with power and data connections and the experiment modules, one of which is accommodated inside the nosecone and is ejectable.

\subsubsection{Flight Details}
\begin{figure}[!bth]
\includegraphics[width=0.5\textwidth]{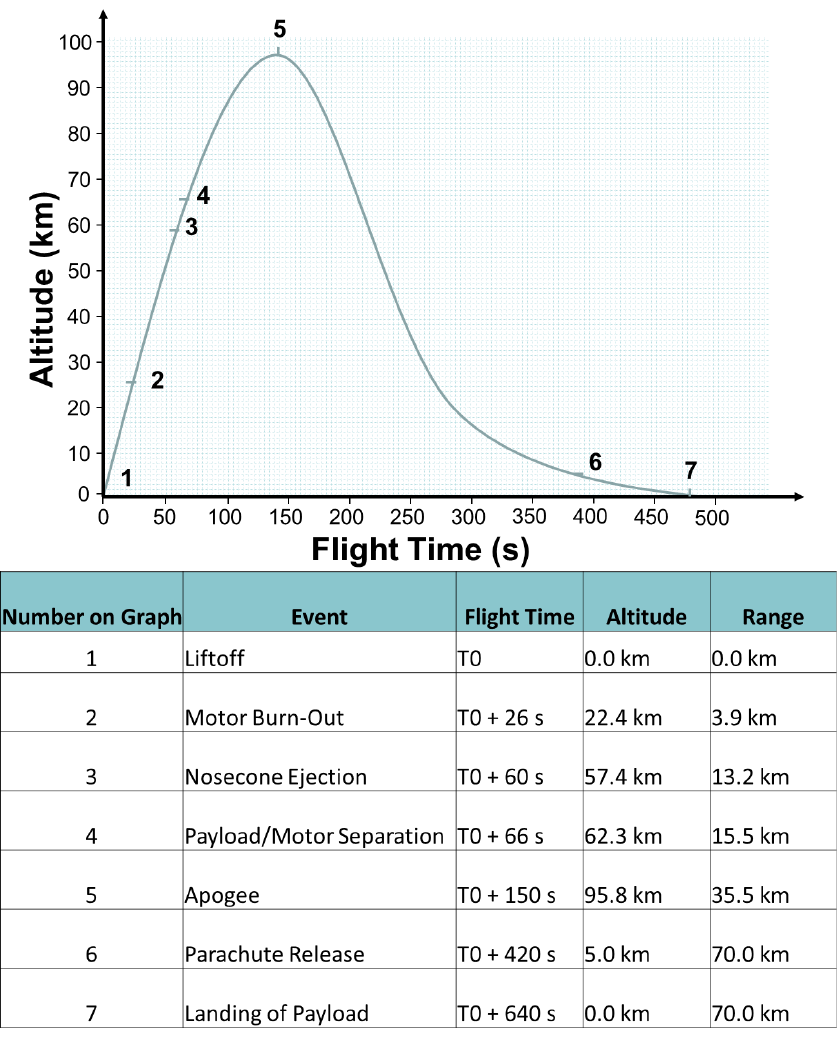}
\caption{A typical REXUS flight profile, i.e. altitude vs. flight time, with flight events numbered (top) and the corresponding event table (bottom) \cite{rexus_manual}. The reduced-gravity phase is the parabola-shaped part of the curve around apogee.}
\label{f:flight_events}
\end{figure}

During a nominal REXUS flight, the rocket motor has burnt out after about 26 s and is being separated from the payload at $\sim$70 s, when the ballistic, i.e., the reduced-gravity flight starts. The low-gravity phase lasts between 150 and 180 s, depending on payload mass (Figure \ref{f:flight_events}). The rocket is spin-stabilized during launch and is being de-spun thereafter by releasing two masses on strings (yo-yo) after the motor burn-out and the nosecone experiment ejection. Apogee is reached at around 90 km altitude about 150 s after liftoff. The payload touches ground again after a flight time of about 640 s, decelerated by the recovery module parachute. Landing occurs inside a range of $\sim$70 km with a nominal velocity of around 8 m/s \cite{rexus_manual}.

\subsubsection{Rocket Interfaces to the Experiments}
For flying onboard the REXUS rocket, experiments have to be designed according to the REXUS User Manual \cite{rexus_manual}. The experiment module available for integrating the SPACE experiment was made of a 14$\verb+"+$ (355.6 mm) diameter, 4 mm thick, 220 mm long aluminum cylinder and an aluminum bulkhead to accommodate most of the experimental components.

The service module of the rocket provides power and communication (telemetry and command) as well as control wires. For each experiment, the available voltage lies between 24 and 36 V and the average electric current should be kept under 1 A, with peak currents of 3 A possible. Telemetry and command can both be implemented via an RS-422 interface. In addition, the service module provides three control lines: the Liftoff (LO), Start-Of-Experiment (SOE) and Start-Of-Data-Storage (SODS) signals can be implemented into the flight sequence of the rocket's onboard software and can be used by each experiment as needed. SPACE used the LO and SOE signals to switch on experiment components automatically during the flight.

\subsection{Launch and Flight Requirements}
In addition to the requirements induced by interfacing the rocket itself, an experiment built for flying on REXUS must also comply to the different launch and flight conditions:
\begin{itemize}
\item Temperature: the launch site being located in northern Sweden, near the city of Kiruna, outside temperatures can reach values down to -40$\celsius$, while the experiment module's skin can heat up to +70$\celsius$ during launch. Experiment components have to be functional (and hence tested) over that range of temperatures. Electronic components in particular, must reliably operate when switched on at low temperatures due to long waiting times of the rocket on the launch rail.
\item Durability: before the launch actually takes place, several tests and countdown simulations are being performed with the hardware being outside on the launch rail. This means that experiment components have to be able to withstand several power cycles at launch site conditions without affecting the experiment run during the actual flight.
\item Late access: once the rocket is assembled, no access to the experiment hardware is granted. If late access is required, an umbilical or hatches have to be included in the experiment design.
\item Loads: a peak acceleration of about 20 g is reached during launch. On top of that, the rocket  and rotates at 3 to 4 Hz inducing corresponding centrifugal forces on the payload. Vibrational loads also have to be considered and the hardware tested according to the requirements \cite{rexus_manual}.
\end{itemize}

\subsection{The SPACE Experiment}
\label{s:SPACE}
The objective of the SPACE experiment is to investigate the collision behavior of submillimeter-sized dust aggregates at low impact velocities. Thus, the hardware is responsible for the generation of the required collisions among the dust aggregates and the software manages the recording of the high-speed video data of the dust-aggregate ensemble throughout the reduced-gravity phase. To achieve this, the experimental setup holds dust aggregates in evacuated glass containers that are back-lit by an LED array and allows for recording the motion of these particles with a high-speed camera.

\begin{figure}[!bth]
\begin{center}
\includegraphics[width=0.45\textwidth]{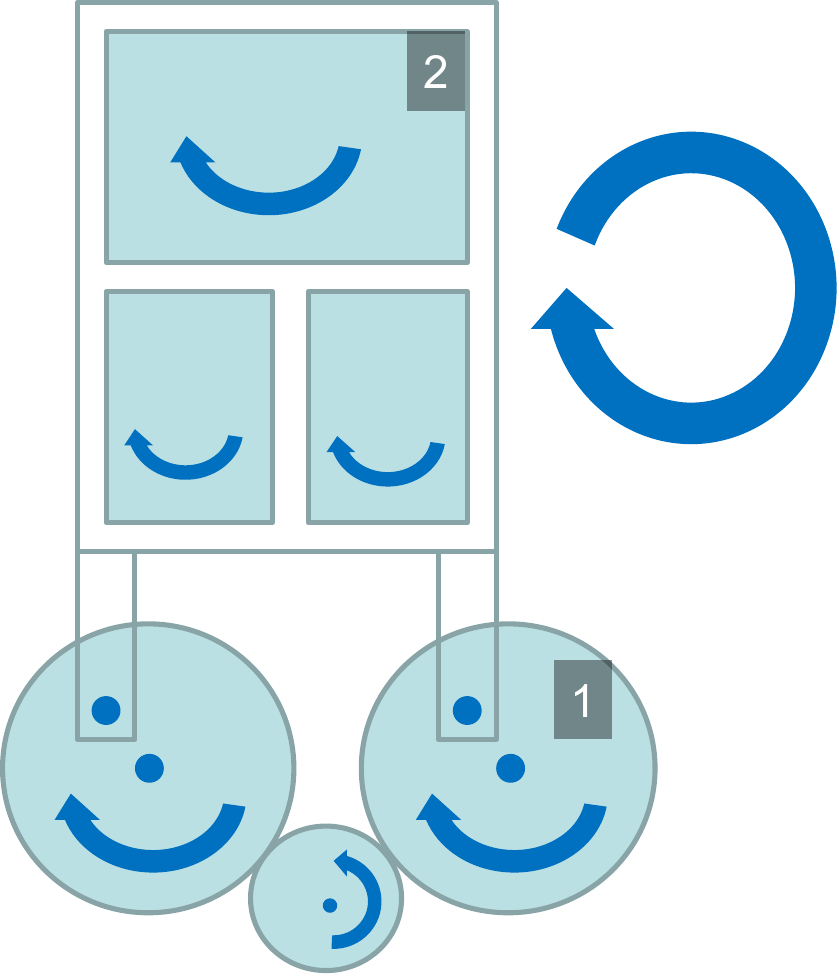}
\caption{Schematics of the two-dimensional shaking mechanism. The rotation of the main cogwheels (1) leads to a circular motion of the particle containers (2). }
\label{f:intern}
\end{center}
\end{figure}

The experimental set-up encompasses three particle containers allowing for some variation of particle properties. To ensure that the collision behavior of the particles is not influenced by the presence of gas, the particle containers are placed inside a vacuum chamber which is evacuated during the experiment run. As the residual gas drag and the residual spin acceleration of the rocket might influence the free motion of the dust aggregates, a two-dimensional shaking mechanism was added, which is being powered by a motor and agitates the particle containers in a rotary fashion (see Figure \ref{f:intern}).

\subsubsection{Particle Containers}
\begin{figure}[!bth]
\begin{center}
\includegraphics[width=0.45\textwidth]{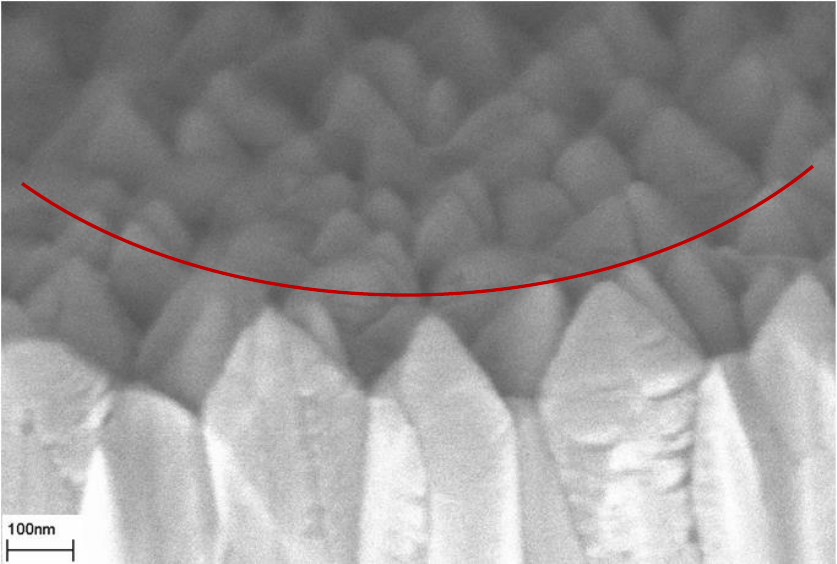}
\caption{SEM image of the nano-particle glass coating on the inner walls of the SPACE glass cells. The red arch represents the typical size of a dust monomer-particle (as part of a dust aggregate) colliding with the wall.}
\label{f:coating}
\end{center}
\end{figure}

The particle containers consist of three glass boxes holding the dust aggregates throughout the experiment duration. The walls of these boxes are flat and made of Borofloat 33 glass pieces except for the two internal walls, which are aluminum surfaces, to investigate the difference in the sticking behavior of the dust aggregates to glass and aluminum targets. The inner glass walls have been coated with a special anti-adhesive nano-particle layer provided by the Fraunhofer Institute for Surface Engineering and Thin Films in Braunschweig. This coating consists of nanometer-sized tips intended to prevent the dust aggregates from sticking to the walls (see Figure \ref{f:coating}).

Two of the dust-aggregate containers are equal-sized with dimensions of 11$\times$10$\times$15 mm$^3$ and the third glass container is bigger with dimensions of 24$\times$10$\times$15 mm$^3$.

\begin{figure}[!bth]
\begin{center}
\includegraphics[width=0.45\textwidth]{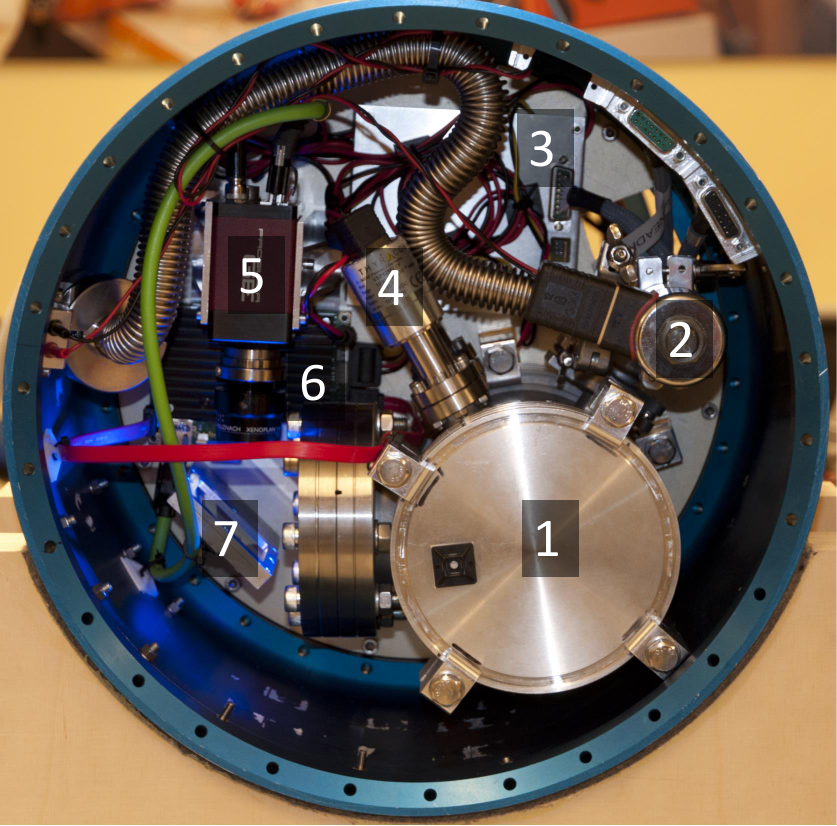}
\caption{The SPACE hardware integrated in its REXUS experiment module. Vacuum chamber (1), vacuum valve (2), electronics board (3), pressure sensor (4), high-speed camera (5), onboard computer (6), and mirror (7).}
\label{f:SPACE}
\end{center}
\end{figure}

\begin{figure}[!bth]
\begin{center}
\includegraphics[width=0.4\textwidth]{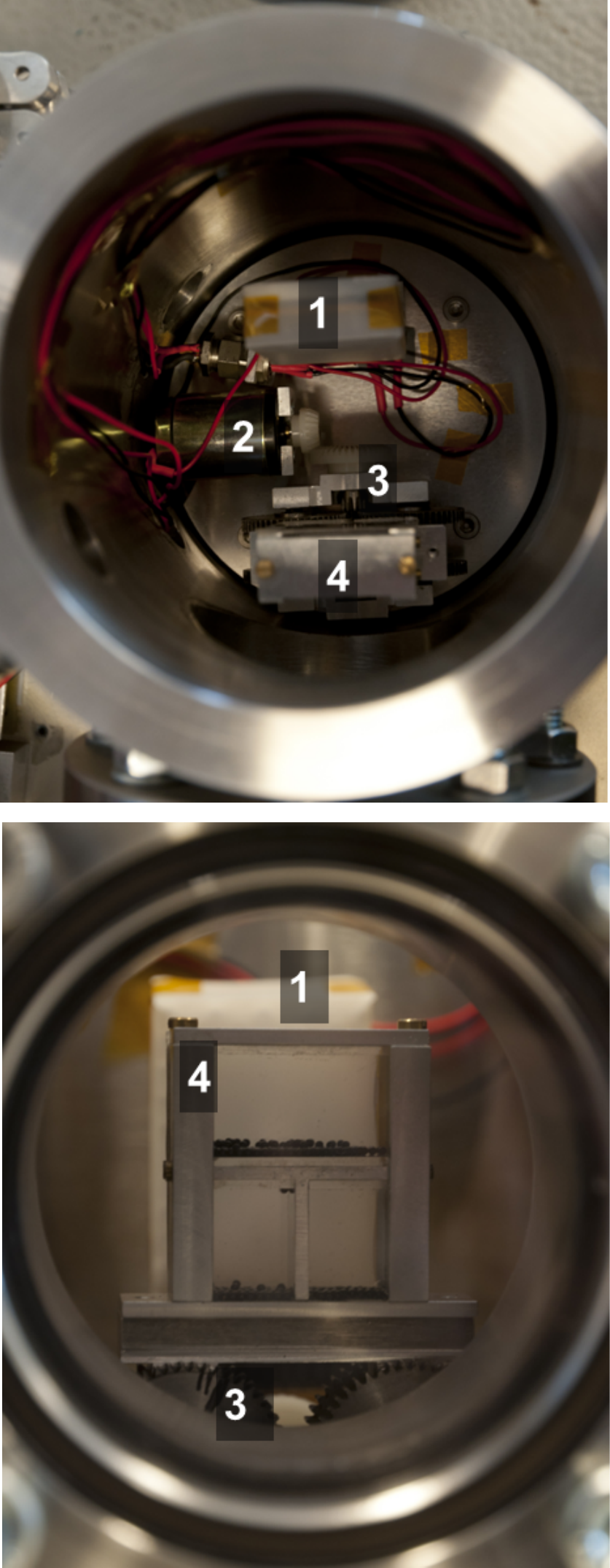}
\caption{The SPACE components inside the vacuum chamber, seen from the top flange (top) and through the chamber's view port (bottom), respectively. LED array covered with diffusion paper (1), motor (2), shaking mechanism (3), and particle containers (4). On these pictures, the particle containers are filled with test particles used to adjust the optical settings of the experiment.}
\label{f:inside_chamber}
\end{center}
\end{figure}

The walls of the particle containers are not perfectly contacting one another, leaving slits narrow enough to allow for the evacuation of air while keeping the dust aggregates inside their cells. To be able to gather data on particle collisions free from outside influences, the experiment was performed under vacuum conditions (at pressures below 10$^{-4}$ mbar). Therefore, the particle containers were built into a vacuum chamber that was evacuated through a Balzers EVC 110M electronic valve. The motor, shaking mechanism and LED array were built into the vacuum chamber (Figures \ref{f:SPACE} and \ref{f:inside_chamber}).

\begin{figure}[!bth]
\includegraphics[width=0.5\textwidth]{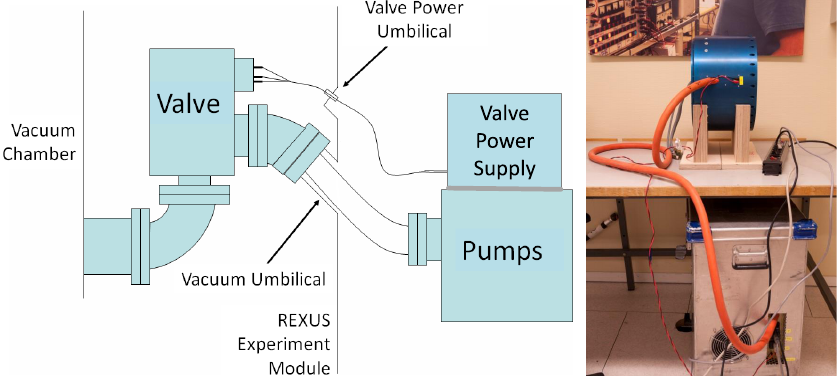}
\caption{The SPACE vacuum umbilical. On the left, a schematic illustration of the umbilical concept is shown. On the right, a picture of the SPACE experiment being evacuated through the umbilical is shown. The SPACE hardware in its blue REXUS module can be seen on the table, whereas the Zarges box is underneath, containing the pre-vacuum and the turbomolecular pumps as well as the valve power supply. Vacuum chamber and pumps are connected via a vacuum-tight rubber hose, specified for cold temperatures down to -40$\celsius$ and power cables.}
\label{f:umbilical}
\end{figure}

To ensure the best vacuum conditions possible during flight, and because the pre-vacuum and turbomolecular pumps could not be integrated into the rocket (for obvious weight and dimension reasons), the outside shell of the SPACE module was outfitted with an umbilical allowing for external evacuation of the vacuum chamber and powering of the vacuum valve. The two pumps were assembled into a Zarges box that could be attached to the launch rail of the rocket. This way, evacuating the chamber was possible until 120 s before launch (Figure \ref{f:umbilical}).

\subsubsection{Shaking of the Dust Aggregates}
\label{p:shaking}
There are several rationales for shaking the particle containers during the experimental run of SPACE. First, shaking provides for a uniform distribution of the dust aggregates in the glass cells. This is not only required because the launch accelerations agglomerate the dust on one side of the particle containers, but also compensates for disturbances of the weightlessness (due to atmospheric drag and residual rocket rotation) throughout the flight phase. During the flight with a REXUS rocket, the achieved microgravity level is affected by two factors. One is the drag produced by the residual atmosphere at the altitudes of the rocket's parabola. For REXUS 12, the apogee of the rocket's trajectory was at about 86 km. For the time between 65 and 275 s after liftoff, the camera recorded the motion and collision of dust aggregates under reduced gravity-conditions. During this period, REXUS 12 was above 60 km in altitude and the residual atmospheric drag resulted in accelerations of less than 10$^{-4}$g in the direction of the roll rotational axis of the rocket. The second factor is the residual spin of the rocket after yo-yo de-spin. As the rocket is spin-stabilized during launch, two masses attached to strings are released after motor burn-out. The spin cannot completely be eliminated and for REXUS 12 the module was still rotating at about 11$\degree$/s during the experimental run. As the particle containers were placed at about 40 mm from the rocket's roll rotation axis, the residual spin resulted in an outward radial acceleration of 1.45$\times$10$^{-3}$g.

Hence, if left freely floating, the dust particles would have the tendency to accumulate in one corner of the particle container under the combined effects of residual centrifugal forces and atmospheric drag. To avoid this behavior, the experiment was outfitted with a shaking mechanism agitating the particles along two directions by applying  a circular motion to the glass containers (see Figure \ref{f:intern}). The agitation mechanism was realized with a motor and several cog-wheels under the frame of the containers (Figure \ref{f:intern} and numbers 2 and 3 in Figure \ref{f:inside_chamber}). 

\begin{figure}[!bth]
\begin{center}
\includegraphics[width=0.5\textwidth]{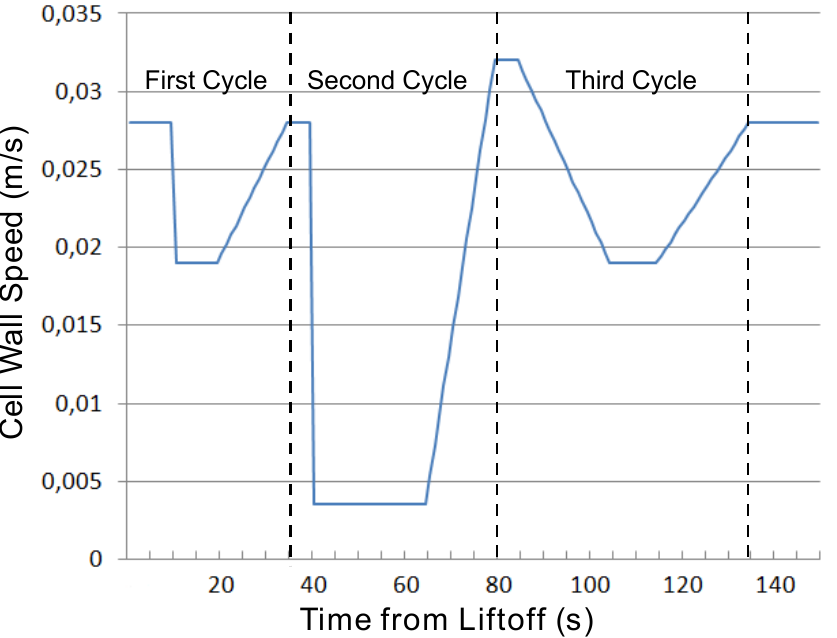}
\caption{Profile of the velocity of the glass-container walls during the REXUS 12 SPACE experimental run.}
\label{f:motor_profile}
\end{center}
\end{figure}

The shaking motion of the particle containers was not only a counter-measure to the residual accelerations, it moreover provided a way of adjusting the internal kinetic energy of the many-particle system. By choosing  a specific shaking profile, the collision speeds between particles can be adjusted and thresholds for growth or disruption of dust aggregates can be investigated. For the SPACE experiment onboard the REXUS 12 flight, the shaking profile was divided in three distinctive sequences (see Figure \ref{f:motor_profile}).
\begin{itemize}
\item First cycle: after an initial shake-up of 10 s duration at full speed (100\% of the motor's nominal voltage is applied) to disintegrate clumps formed during launch, the motor voltage was reduced to half its nominal value for 10 s and ramped up to 100\% again within 15 s. It should be mentioned that, due to the non-linearity of the motor drive, the rotation speed decreased less than 50\%.
\item Second cycle: after an initial shake-up of 5 s duration to disrupt agglomerates formed in the previous cycle, the voltage applied to the motor was reduced to 20\% of its nominal value for a duration of 25 s. This corresponds to the flight phase around the apogee of the rocket's trajectory. The minimum acceleration transferred from the particle container walls to the aggregates during this phase is on the order of 10$^{-3}$ g. The motor voltage was then ramped up to 125\% of its nominal value within 15 s to observe the aggregates fragmenting over a large range of speeds.
\item Third cycle: after an initial shake up of 5 s duration to disrupt agglomerates of the previous cycle, the motor input voltage was ramped down to 50\% of its nominal value within 20 s to observe the agglomeration of particles over a large range of speeds. The voltage was then kept at 50\% for 10 s and ramped up to 100\% again within 20 s.
\item After the last cycle: the motor was kept running for 15 s.
\end{itemize}

\subsubsection{Observation of the Dust Aggregates}
The particle containers are illuminated by an LED array positioned behind the glass cells. The LED array is composed of 86 blue LEDs, each 3 mm in diameter, distributed over 9 rows. Blue LEDs were used because the camera sensitivity is best at this specific wavelength, compared to white light or other colors. To obtain a uniform back illumination over the entire field of view of the camera, the LED array and the backside of the particle containers are covered with a sheet of diffusion paper.

\begin{figure}[!bth]
\begin{center}
\includegraphics[width=0.475\textwidth]{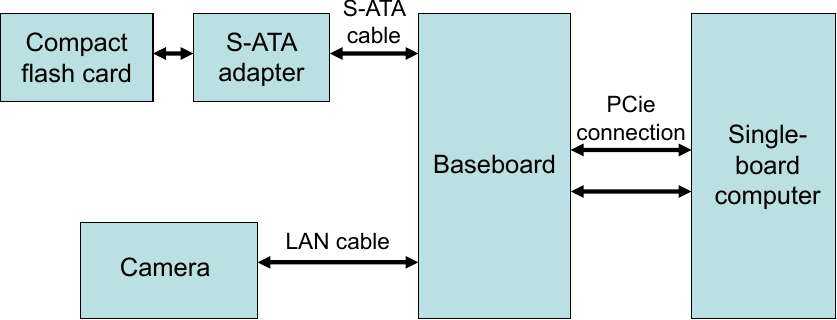}
\caption{Schematics of the SPACE image acquisition concept.}
\label{f:acquisition}
\end{center}
\end{figure}

The particle collisions during the experimental run were recorded by an Allied Vision Technologies Prosilica GE680 high speed camera at a continuous rate of 170 frames per second and a resolution of 640$\times$480 pixels. This camera was chosen as a compromise between the size and weight restrictions for REXUS experiments and the imaging performance. The Prosilica GE680 camera has no internal control nor recording abilities and must be commanded by an external computer. For this purpose, a combination of a Toradex single-board computer Robin Z530 L and its interface board Daisy Pico-ITX was built into the experiment. These components were chosen because the Robin computer comes with an RTL 8111D Ethernet controller capable of handling the Jumbo packets provided by the Prosilica camera at high frame rates. The Daisy carrier board allows for the connection between the Robin computer and the camera via an Ethernet cable. Linux Ubuntu Server 11.0 was used as an operating system for its ability to implement all the required interfaces while using only very little of the computer's memory space. The C++ software controlling the camera was responsible for configuring the camera, starting the frame acquisition and recording the streamed frames to an external compact flash card (Figure \ref{f:acquisition}).\

\begin{figure}[!bth]
\begin{center}
\includegraphics[width=0.475\textwidth]{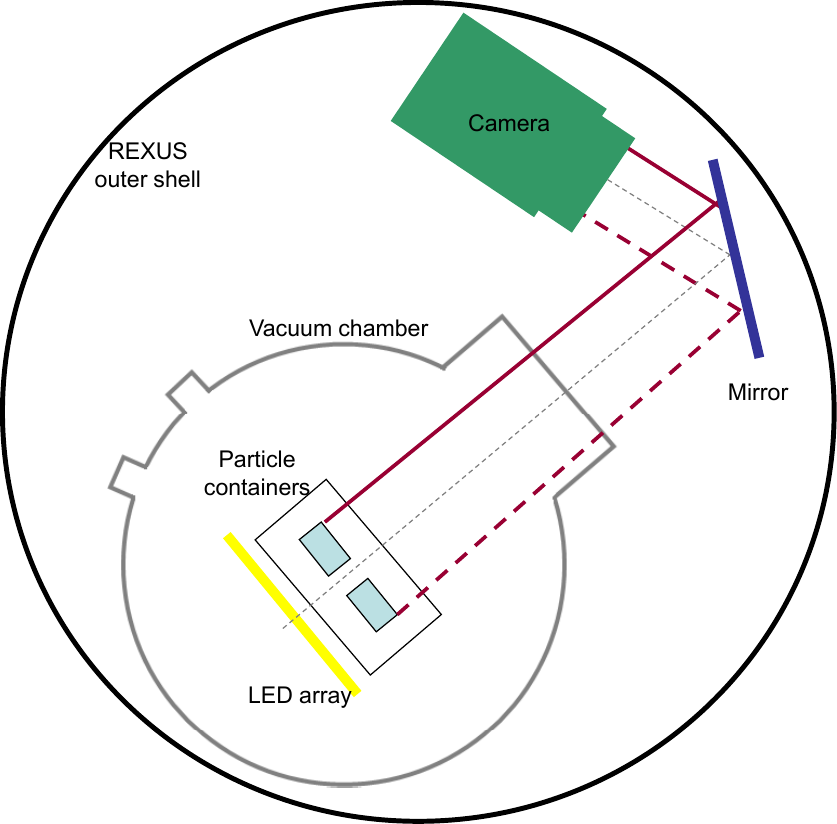}
\caption{Schematics of the optical path in the SPACE experiment.}
\label{f:optical_path}
\end{center}
\end{figure}

The optical path between the particle containers and the high-speed camera had to be adapted to the module size of the REXUS rocket. An additional mirror between the window of the vacuum chamber and the camera lens folded the optical path and allowed for fitting the experimental hardware into the required dimensions of the REXUS rocket module (Figure \ref{f:optical_path}).

\subsubsection{Autonomous Run of the Experimental Procedures}
As the duration of a REXUS flight is quite short (with about three minutes reduced-gravity time), it was decided to run the experiment fully autonomously, using only the control wires of the rocket's service module instead of up-linking commands. To that purpose, an electronics board was designed and manufactured that was responsible for executing the  pre-defined experimental procedures along a timeline. An ATMega32 micro-controller switched the experimental components on and off via solid state relays (AQV 252G). This was done either upon receiving a signal from the service module in flight configuration, or commands from the ground station in test configuration. The micro-controller also produced health and status telemetry while the experiment was powered. The onboard computer and electronics board are automatically switched on and kept in stand-by mode when the experiment is powered. The LO signal is used to enter the flight mode. Upon reception of the SOE signal, the micro-controller switches on the camera, the illumination and runs through the shaking profile of the motor as described above and in Figure \ref{f:motor_profile}. Once the sequence has terminated, motor, illumination and camera are switched off. Un-powering of the experiment is performed directly by the rocket's service module, ensuring that the recorded imaging and housekeeping data cannot be overwritten or erased during descent and landing (Table \ref{t:timeline}).

At the same time that the experiment timeline is being worked through, the micro-controller also switches on the pressure sensor and records its readings on its own EEPROM memory.

\subsubsection{Data Rate and Data Volume}
During a typical experimental run, the camera produces 170 frames per second, each one containing 640$\times$480 pixels at 8 bit grayscale values. This amounts to a data rate of about 61 MB/s flowing from the camera to the onboard computer and being recorded to an external flash card. To achieve this high data rate, the camera is connected to the onboard computer via an Ethernet cable and streams Jumbo packets. On the recording side, a 600x Transcend compact flash card is connected to the Daisy/Robin unit via a S-ATA cable and the associated flashcard to S-ATA adapter (Figure \ref{f:acquisition}). An image-recording sequence of about 210 s duration produces about 11 GB of data, which can be stored on the 16 GB compact flash card in packages of $\sim$ 800 MB each.

\section{The Experiment Run of the SPACE Experiment onboard REXUS 12}
\subsection{Dust Samples}
\begin{figure}[!bth]
\begin{center}
\includegraphics[width=0.475\textwidth]{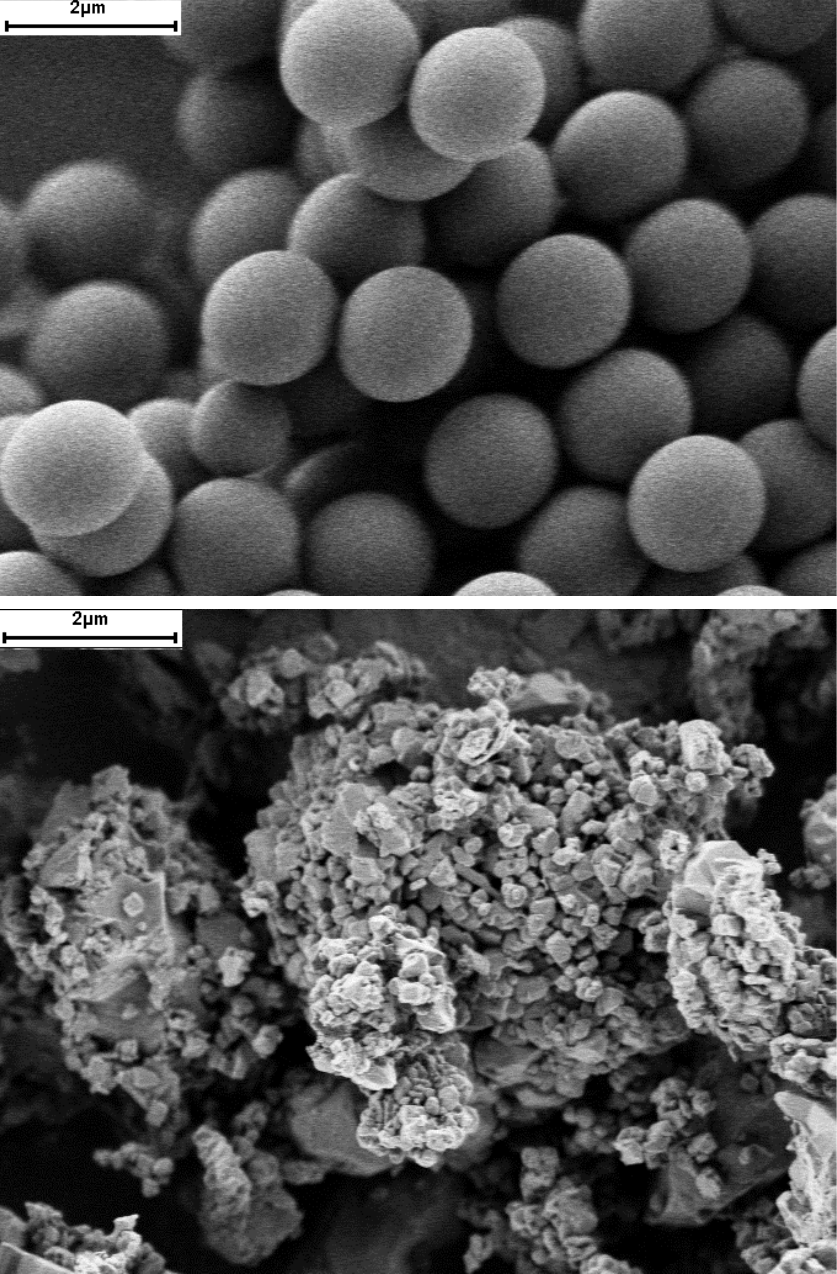}
\caption{SEM pictures of monodisperse (top) and polydisperse (bottom) SiO$_2$ dust particles.}
\label{f:dust}
\end{center}
\end{figure}

The dust samples used in the SPACE experiment onboard REXUS 12 were composed of SiO$_2$ aggregates of two different kinds and size distributions. We used monodisperse, spherical SiO$_2$ particles as well as aggregates constituting of polydisperse, irregular SiO$_2$ grains (Figure \ref{f:dust}).

\begin{figure}[!bth]
\begin{center}
\includegraphics[width=0.475\textwidth]{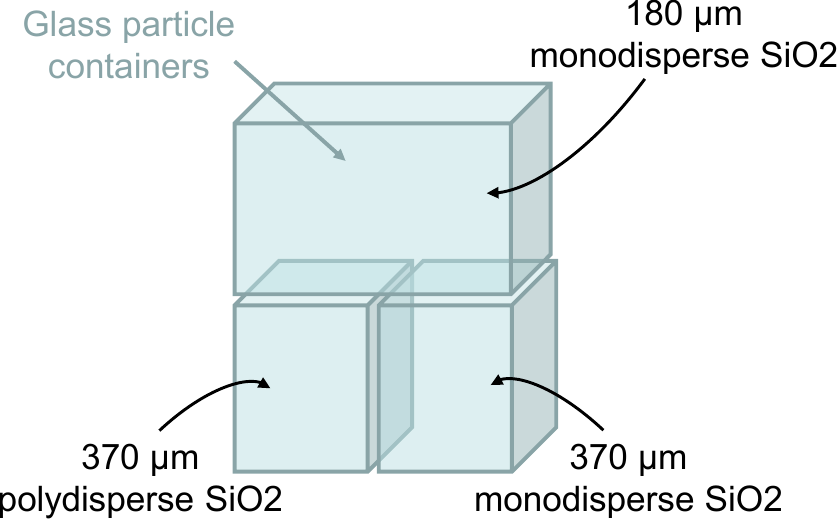}
\caption{Arrangement of the SiO$_2$ dust types and sizes in the SPACE containers during the REXUS 12 flight.}
\label{f:container_filling}
\end{center}
\end{figure}

The monodisperse SiO$_2$ aggregates were sieved into two different size categories with mean aggregate diameters of 180 $\mu$m and and 370 $\mu$m, respectively. Polydisperse SiO$_2$ aggregates were sieved to a mean diameter of 370 $\mu$m only. The larger particle container (with a size of 24$\times$10$\times$15 mm$^3$) was filled with the 180 $\mu$m-sized monodisperse aggregates, one of the smaller containers (with a size of 11$\times$10$\times$15 mm$^3$) with 370 $\mu$m-sized monodisperse aggregates, and the second small container with 370 $\mu$m-sized polydisperse aggregates, respectively (Figure \ref{f:container_filling}).

\subsection{The REXUS 12 Flight}
On March 19, 2012, at 14:05 UTC, REXUS 12 was successfully launched with the SPACE experiment aboard. After motor burn-out and separation at 26 s and 77 s, respectively, the rocket reached apogee at around 86 km altitude after 140 s into the flight. Due to technical issues, however, the parachute did not deploy as expected and the payload impacted the ground with a much higher velocity than nominal. The rocket and its payloads were nonetheless recovered and the individual modules could be handed back to the experimenters. The SPACE experiment was powered 600 s before liftoff, received the LO and SOE signals at 0 and 65 s after launch, respectively, and ran its complete experimental sequence before being switched off by the rocket timeline at 330 s after liftoff (Table \ref{t:timeline}). The whole time it was powered, SPACE delivered nominal health and status telemetry indicating a nominal run through the internal experiment timeline. Upon recovery, most of the experimental hardware was damaged, due to the hard landing, but the compact flash card, which had been safely built in below the onboard computer and was holding the scientific data, could be retrieved intact. The electronics board including the micro-controller could also be recovered functional, delivering the flight pressure data of the evacuated chamber.

\begin{table}[htbp]
\caption{The REXUS 12 and SPACE internal timeline.}
\begin{tabular}{|l|l|l|}
\hline

\textbf{Time to}&\textbf{REXUS event}& \textbf{SPACE internal} \\
\textbf{launch [s]}&&\textbf{event}\\ \hline

T-600&Experiments&Main power on by\\
&Power On&the REXUS Service\\
&&Module. The onboard\\
&&computer, flash card\\
&&and electronics board\\
&&are powered.\\ \hline
T-120&REXUS Service&End of SPACE\\
&Module switch&vacuum chamber\\
&to internal power&evacuation. \\ \hline
T+0&Liftoff&\\ \hline
T+0&LO Signal to all&SPACE enters\\
& Experiments&the flight mode.\\ \hline
T+26&Burnout of the&\\
&rocket motor&\\
&(Improved Orion)&\\ \hline
T+65&SPACE SOE Signal&The micro-controller\\
&&switches on the\\
&&LED array, camera\\
&&and motor.\\ \hline
T+70&Yo-yo Despin&\\ \hline
T+74&Nosecone Ejection&\\ \hline
T+77&Motor Separation&\\ \hline
T+95&&The micro-controller\\
&&starts the shaking\\
&&sequence for the\\
&&motor.\\ \hline
T+142&Apogee&\\ \hline
T+330&SPACE Power Off&All experimental\\
&&components are off.\\ \hline
T+600&Experiments&\\
&Power Off&\\ \hline
\end{tabular}
\label{t:timeline}
\end{table}

\subsection{Data Obtained During the REXUS Flight of SPACE}
The high-speed imaging data retrieved from the compact flash card turned out to be of the expected quality and can be used for the analysis of the collision behavior of the dust aggregates as intended. Only a few discrepancies to a completely nominal experimental run occurred, which, however, do not compromise the scientific analysis of the SPACE experiment:
\begin{itemize}
\item The camera run-up lasted about three times longer than during ground tests; instead of the usual 10 s from power-up to start of frame acquisition on ground, about 30 s were required during flight. The reason for this anomaly could not be determined.
\item Some frames of the high-speed camera image sequences were lost during the first 10 s of the data recording. This behavior had already been observed during ground tests and had been taken into account in the experiment timeline by implementing a longer recording time.
\item The sticking efficiency of the submillimeter-sized dust aggregates and the anti-adhesive glass walls of containers was much higher than expected and tested before during drop-tower experiments. Especially during slow shaking phases, dust aggregates tended to form clusters on the glass walls rather than colliding with one another while flying freely, as done in the almost perfect microgravity environment of the drop tower.
\end{itemize}

\begin{figure}[!bth]
\begin{center}
\includegraphics[width=0.475\textwidth]{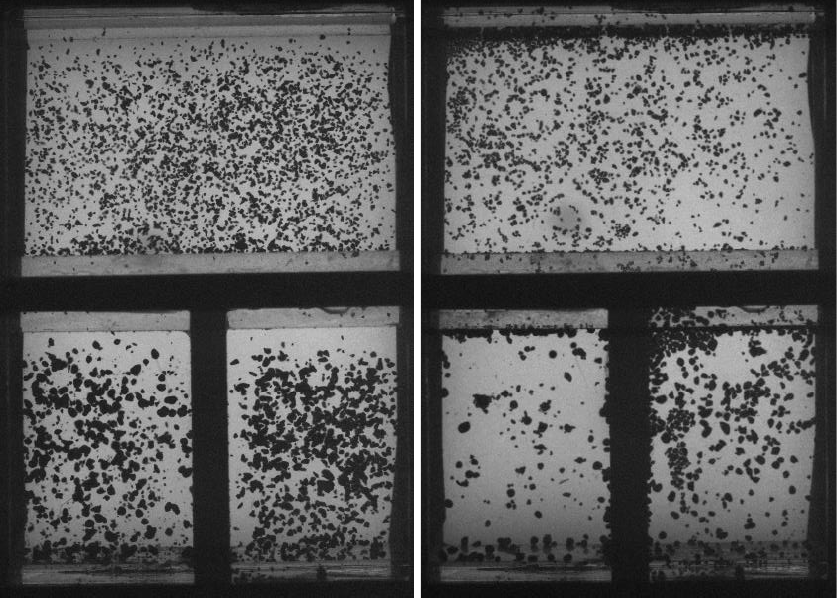}
\caption{Two frames of the high-speed camera, recorded during the REXUS 12 flight. On the left, an image at the very beginning of the experimental run is shown. On the right, the three glass containers can be seen towards the end of the reduced-gravity phase. An accumulation of particles into the top-left corners of the particle containers can clearly be observed towards the end of the experimental run.}
\label{f:frames}
\end{center}
\end{figure}

These points lead to a reduction of the usable amount of data from the expected 11 GB to about 8 GB, which is still a substantial quantity of material for the purpose of the intended analysis (see Figure \ref{f:frames} for an example).

\section{Data Analysis}
The full analysis of the data obtained by the SPACE experiment onboard the REXUS 12 flight is still ongoing and will be combined with the data sampled by the experiment during its hardware test in the Bremen drop tower in August 2011.

\subsection{Analysis Methods}
\begin{figure}[!bth]
\begin{center}
\includegraphics[width=0.475\textwidth]{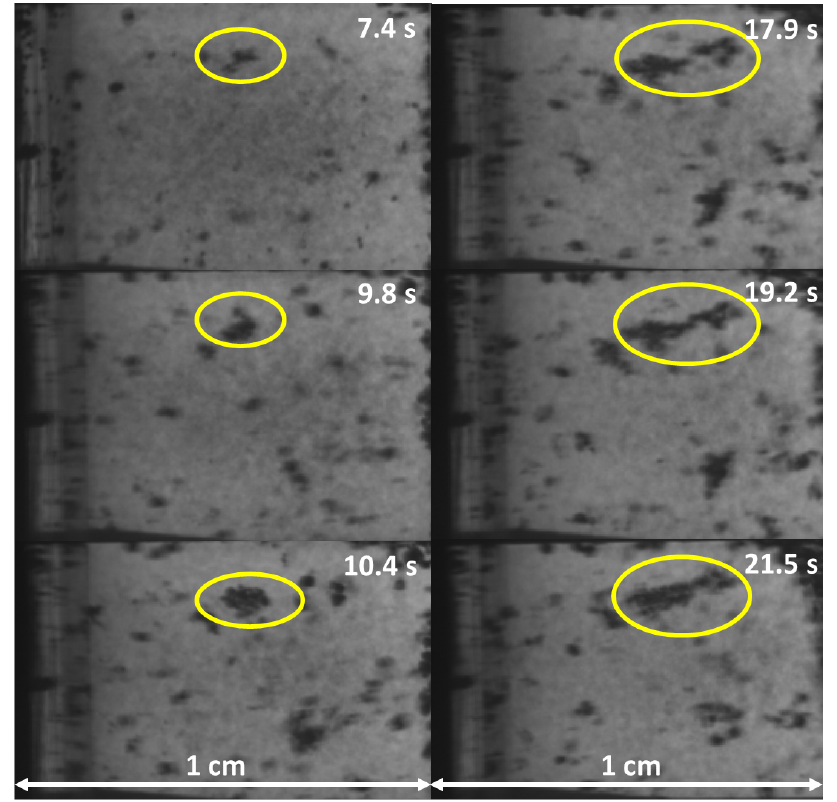}
\caption{Example of an aggregate growing in one of the particle containers of the SPACE experiment during the REXUS 12 flight (encircled). Each of the six images has been averaged over 201 frames (just over 1 s). Time is counted from the start of recording by the camera.}
\label{f:boxcar}
\end{center}
\end{figure}

The high-speed imaging data recorded during the experimental run of SPACE comes in form of 8 bit grayscale bitmap frames (see Figure \ref{f:frames}). These pictures are analyzed with self-written software using IDL (Interactive Data Language). Part of the basic image processing is for example the elimination of the rotational motion of the frames due to the circular shaking of the glass containers and the correction of remaining back-illumination irregularities. The real-time shaking frequency of the glass cells and the particle velocities can directly be determined by the known frame rate of the camera. As the many-particle systems in the SPACE experiments are optically quite dense, the growth rate of aggregates can best be investigated by averaging a certain number of pictures before and after each frame. An example of this procedure is shown in Figure \ref{f:boxcar}.

\begin{figure}[!bth]
\begin{center}
\includegraphics[width=0.475\textwidth]{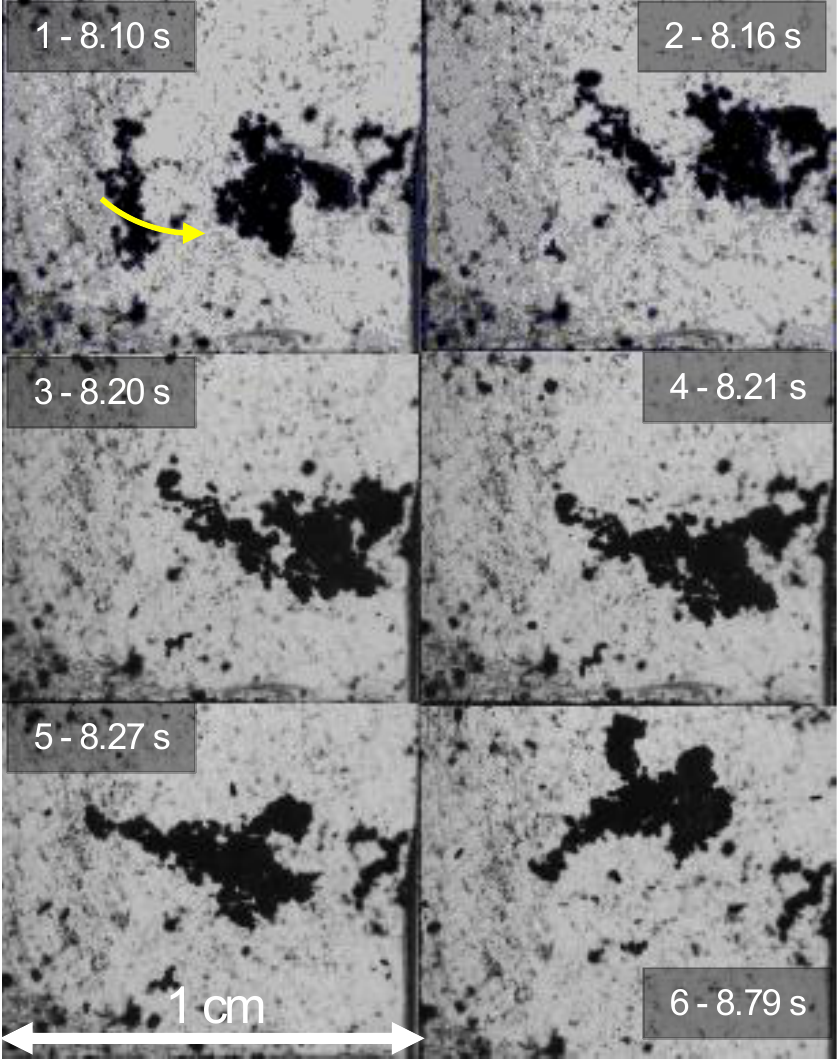}
\caption{Two dust aggregates colliding and sticking in one of the particle containers of the SPACE experiment during a hardware test at the Bremen drop tower. The relative velocity of the aggregates is 1.61 cm/s. Time is counted from the recording start of the camera.}
\label{f:collision}
\end{center}
\end{figure}

In individual cases, the SPACE hardware allows the direct observation of collisions between two dust aggregates, as shown in Figure \ref{f:collision}, recorded during a drop-tower test run. The dust aggregates can then be tracked along several frames and their collision velocity can be determined.

\subsection{Expected Scientific Results}
By observing the growth rate of dust aggregates in the SPACE particle containers during the REXUS 12 flight, and by relating them to the shaking frequencies induced by the motor, a threshold velocity for sticking of submillimeter-sized dust aggregates can be determined. In the same way, an aggregate fragmentation limit can be investigated. In addition, the structure of the growing dust aggregates and its role in the agglomerate growth can also be determined. Finally, the sticking or bouncing outcome of individual collisions between dust aggregates can provide data points to the dust collision model presented in section \ref{s:intro}. The full results of the SPACE data analysis will be the subject of a dedicated paper.

\section{Lessons Learned}
\subsection{Hardware}
If the SPACE experiment were to be re-built, the first recommendation would be to either choose a camera displaying a higher level of autonomy (i.e. internal memory) or using an onboard computer with a bigger RAM space. The Robin Z530 L single-board computer possesses a RAM of 1 GB, which it uses as buffer for incoming frames from the camera while it writes them on the external flash card. For higher acquisition rates, this buffer space is insufficient and the frame streaming starts experiencing severe data losses as the RAM overloads. This could be made up for by allocating additional swap space on available hard disk memory. However, the maximal possible frame rate for the camera (205 fps) could not be achieved and a compromise at an acceptable rate of 170 fps, at which no considerable data loss occurred, had to be consented to.

\subsection{Scientific Data}
The experience of the REXUS 12 flight has shown that the dust aggregates used in the SPACE experiment possess a very high sticking efficiency with the glass walls of the particle containers, even though these were actually coated with a nano-particle anti-adhesive layer. Hence, the shaking frequency of the particle containers should be adjusted accordingly to keep the particles as much as possible free-floating in the inner container volume and therewith optimizing the quantity of usable data among the recorded frames. The acceleration level due to residual atmospheric drag and rocket spin can also clearly be observed in the experimental data by the accumulation of particles in a specific corner of the particle containers (see Figure \ref{f:frames}). Having now gathered experience on the strength and direction of this perturbation, the shaking profile of a future experiment run could also be adapted accordingly.

\begin{acknowledgements}
We thank the REXUS/BEXUS project of the Deutsches Zentrum f{\"{u}}r Luft- und Raumfahrt (DLR) for the flight on the REXUS 12 rocket and for their contribution towards hardware expenses. This work was supported by the ICAPS (Interaction in Cosmic and Atmospheric Particle Systems) project of DLR (grant 50WM0936) and a fellowship from the International Max Planck Research School on Physical Processes in the Solar System and Beyond (IMPRS). We also thank Oliver Werner from the Fraunhofer Institute for Surface Engineering and Thin Films of Braunschweig for the anti-adhesive glass coating of the particle containers.
\end{acknowledgements}


%

\end{document}